\documentclass[modern]{aastex63}

\usepackage{xspace}
\usepackage{comment}

\newcommand{\mysec}[1]{\S\ref{#1}\xspace}
\newcommand{\myeq}[1]{Equation~\ref{#1}\xspace}
\newcommand{\myfig}[1]{Figure~\ref{#1}\xspace}
\newcommand{\mytab}[1]{Table~\ref{#1}\xspace}

\newcommand{\enterprise}{\textsc{Enterprise}\xspace}
\newcommand{\extensions}{\texttt{enterprise\_extensions}\xspace}
\newcommand{\tempo}{\textsc{Tempo}\xspace}
\newcommand{\tempotwo}{\textsc{Tempo2}\xspace}
\newcommand{\pint}{\textsc{Pint}\xspace}

\received{\today}
\submitjournal{ApJ}

\shorttitle{Bayesian Solar Wind}
\shortauthors{Hazboun et al.}
\graphicspath{{./}{figures/}}

\begin{document}

\title{Bayesian Solar Wind Modeling with Pulsar Timing Arrays}



\correspondingauthor{Jeffrey S. Hazboun}
\email{hazboun@uw.edu}

\author[0000-0003-2742-3321]{Jeffrey S. Hazboun}
\altaffiliation{NANOGrav Physics Frontiers Center Postdoctoral Fellow}
\affiliation{Physical Sciences Division, University of Washington Bothell, 18115 Campus Way NE, Bothell, WA 98011, USA}
\author[0000-0003-1407-6607]{Joseph Simon}
\affiliation{Department of Astrophysical and Planetary Sciences, University of Colorado, Boulder, CO 80309, USA}
\author[0000-0003-2285-0404]{Dustin R. Madison}
\affiliation{Department of Physics, University of the Pacific, 3601 Pacific Avenue, Stockton, CA 95211, USA}

\author{Zaven Arzoumanian}
\affiliation{X-Ray Astrophysics Laboratory, NASA Goddard Space Flight Center, Code 662, Greenbelt, MD 20771, USA}
\author[0000-0002-1529-5169]{Kathryn Crowter}
\affiliation{Department of Physics and Astronomy, University of British Columbia, 6224 Agricultural Road, Vancouver, BC V6T 1Z1, Canada}
\author[0000-0002-2185-1790]{Megan E. DeCesar}
\affiliation{George Mason University, resident at the Naval Research Laboratory, Washington, DC 20375, USA}
\author[0000-0002-6664-965X]{Paul B. Demorest}
\affiliation{National Radio Astronomy Observatory, 1003 Lopezville Rd., Socorro, NM 87801, USA}
\author[0000-0001-8885-6388]{Timothy Dolch}
\affiliation{Department of Physics, Hillsdale College, 33 E. College Street, Hillsdale, MI 49242, USA}
\affiliation{Eureka Scientific, Inc. 2452 Delmer Street, Suite 100, Oakland, CA 94602-3017}
\author{Justin A. Ellis}
\affiliation{Infinia ML, 202 Rigsbee Avenue, Durham NC, 27701}
\author[0000-0002-2223-1235]{Robert D. Ferdman}
\affiliation{School of Chemistry, University of East Anglia, Norwich, NR4 7TJ, United Kingdom}
\author[0000-0001-7828-7708]{Elizabeth C. Ferrara}
\affiliation{Department of Astronomy, University of Maryland, College Park, MD, 20742, USA}
\affiliation{Center for Exploration and Space Studies (CRESST), NASA/GSFC, Greenbelt, MD 20771, USA}
\affiliation{NASA Goddard Space Flight Center, Greenbelt, MD 20771, USA}
\author[0000-0001-8384-5049]{Emmanuel Fonseca}
\affiliation{Department of Physics and Astronomy, West Virginia University, P.O. Box 6315, Morgantown, WV 26506, USA}
\affiliation{Center for Gravitational Waves and Cosmology, West Virginia University, Chestnut Ridge Research Building, Morgantown, WV 26505, USA}
\author[0000-0001-8158-683X]{Peter A. Gentile}
\affiliation{Department of Physics and Astronomy, West Virginia University, P.O. Box 6315, Morgantown, WV 26506, USA}
\affiliation{Center for Gravitational Waves and Cosmology, West Virginia University, Chestnut Ridge Research Building, Morgantown, WV 26505, USA}
\author{Glenn Jones}
\affiliation{Rigetti Computing, Inc., 775 Heinz Ave. Berkeley, CA 94710, USA}
\affiliation{Columbia Astrophysics Laboratory, Columbia University, NY 10027, USA}
\author[0000-0001-6607-3710]{Megan L. Jones}
\affiliation{Center for Gravitation, Cosmology and Astrophysics, Department of Physics, University of Wisconsin-Milwaukee,\\ P.O. Box 413, Milwaukee, WI 53201, USA}
\author[0000-0003-0721-651X]{Michael T. Lam}
\affiliation{School of Physics and Astronomy, Rochester Institute of Technology, Rochester, NY 14623, USA}
\affiliation{Laboratory for Multiwavelength Astrophysics, Rochester Institute of Technology, Rochester, NY 14623, USA}
\author[0000-0002-2034-2986]{Lina Levin}
\affiliation{Jodrell Bank Centre for Astrophysics, School of Physics and Astronomy, The University of Manchester, Manchester M13 9PL, UK}
\author[0000-0003-1301-966X]{Duncan R. Lorimer}
\affiliation{Department of Physics and Astronomy, West Virginia University, P.O. Box 6315, Morgantown, WV 26506, USA}
\affiliation{Center for Gravitational Waves and Cosmology, West Virginia University, Chestnut Ridge Research Building, Morgantown, WV 26505, USA}
\author[0000-0001-5229-7430]{Ryan S. Lynch}
\affiliation{Green Bank Observatory, P.O. Box 2, Green Bank, WV 24944, USA}
\author[0000-0001-7697-7422]{Maura A. McLaughlin}
\affiliation{Department of Physics and Astronomy, West Virginia University, P.O. Box 6315, Morgantown, WV 26506, USA}
\affiliation{Center for Gravitational Waves and Cosmology, West Virginia University, Chestnut Ridge Research Building, Morgantown, WV 26505, USA}
\author[0000-0002-3616-5160]{Cherry Ng}
\affiliation{Dunlap Institute for Astronomy and Astrophysics, University of Toronto, 50 St. George St., Toronto, ON M5S 3H4, Canada}
\author[0000-0002-6709-2566]{David J. Nice}
\affiliation{Department of Physics, Lafayette College, Easton, PA 18042, USA}
\author[0000-0001-5465-2889]{Timothy T. Pennucci}
\affiliation{Institute of Physics, E\"{o}tv\"{o}s Lor\'{a}nd University, P\'{a}zm\'{a}ny P. s. 1/A, 1117 Budapest, Hungary}
\author[0000-0001-5799-9714]{Scott M. Ransom}
\affiliation{National Radio Astronomy Observatory, 520 Edgemont Road, Charlottesville, VA 22903, USA}
\author[0000-0002-5297-5278]{Paul S. Ray}
\affiliation{U.S. Naval Research Laboratory, Washington, DC 20375, USA}
\author[0000-0002-6730-3298]{Ren\'{e}e Spiewak}
\affiliation{Jodrell Bank Centre for Astrophysics, School of Physics and Astronomy, The University of Manchester, Manchester M13 9PL, UK}
\author[0000-0001-9784-8670]{Ingrid H. Stairs}
\affiliation{Department of Physics and Astronomy, University of British Columbia, 6224 Agricultural Road, Vancouver, BC V6T 1Z1, Canada}
\author[0000-0002-7261-594X]{Kevin Stovall}
\affiliation{Department of Physics and Astronomy, University of New Mexico, 210 Yale Blvd NE, Albuquerque, NM 87106, USA}
\author[0000-0002-1075-3837]{Joseph K. Swiggum}
\altaffiliation{NANOGrav Physics Frontiers Center Postdoctoral Fellow}
\affiliation{Department of Physics, Lafayette College, Easton, PA 18042, USA}
\author[0000-0001-5105-4058]{Weiwei Zhu}
\affiliation{National Astronomical Observatories, Chinese Academy of Science, 20A Datun Road, Chaoyang District, Beijing 100012, China}
\collaboration{30}{(The NANOGrav Collaboration)}

\defcitealias{abb+18a}{NG11}
\defcitealias{madison+2019,}{M19}
\defcitealias{tiburzi+2021}{T21}

\begin{abstract}

Using Bayesian analyses we study the solar electron density with the NANOGrav 11-year pulsar timing array (PTA) dataset. Our model of the solar wind is incorporated into a global fit starting from pulse times-of-arrival. We introduce new tools developed for this global fit, including analytic expressions for solar electron column densities and open source models for the solar wind that port into existing PTA software. We perform an {\it ab initio} recovery of various solar wind model parameters. We then demonstrate the richness of information about the solar electron density, $n_E$, that can be gleaned from PTA data, including higher order corrections to the simple $1/r^2$ model associated with a free-streaming wind (which are informative probes of coronal acceleration physics), quarterly binned measurements of $n_E$ and a continuous time-varying model for $n_E$ spanning approximately one solar cycle period. Finally, we discuss the importance of our model for chromatic noise mitigation in gravitational-wave analyses of pulsar timing data and the potential of developing synergies between sophisticated PTA solar electron density models and those developed by the solar physics community.

\end{abstract}


\section{Introduction} \label{sec:intro}

Radio observations of distant astrophysical sources have long been used to study the content and characteristics of the solar wind. For example, \citet{hewish+1967} used observations of scintillation from quasars as a way to probe the structure of the solar wind. The dispersion of pulsed radio emission from pulsars, due to the diffuse ionized medium along the observational line of sight (LOS), has been known from their first radio observations \citep{bell+1968}. The utility of these radio pulses for investigations of electron density was used in \citet{counselman+1970}, only two years after the discovery of pulsars, to measure the solar electron density ($10 \;{\rm cm}^{-3}$ at $1 {\rm au}$) based on a spherically symmetric model for the solar wind. 

Since these early investigations, pulsar astronomers have often included a model for the solar electron density as a part of pulsar ephemerides---see, for example, the three main pulsar timing software packages \tempo, \tempotwo, and \pint \citep{Nice:2015a, tempo2, luo+2021}. In \citet{lommen+2006} and \citet{splaver+2005}, the solar wind signal was shown to be highly covariant with astrometric components of the timing models for individual pulsars, in particular the parallax and sky position, because these parameters contain strong Fourier components at $1/{\rm yr}$ and higher harmonics. 
Beyond the importance of a solar wind model for accuracy in pulsar astronomy, \citet{you+2007} and \citet{you+2012} showed the sensitivity of pulsar data sets to more complex (than the spherically symmetric $1/r^2$ wind) features in the solar electron density\footnote{We present more detail on these models in \mysec{sec:solar_electron}}. 
More recent work \citep{madison+2019,tiburzi+2019,tiburzi+2021} has shown the potential of pulsar timing arrays (PTAs) as independent probes of the solar wind and its behavior as a function of time.

In this paper we introduce new methods, expanding upon those used in \citet{madison+2019} and \citet{tiburzi+2021}, to obtain information about the solar electron density from PTA data. Using a fully Bayesian framework, we show that much more information about the solar wind can be obtained from the same set of pulsar data used in \citet{madison+2019}, the NANOGrav 11-year Data Set \citet[][henceforth referred to as \citetalias{abb+18a}]{abb+18a}. As in these recent studies, it is important that we are using an {\it array} of pulsars. This allows us to separate the variations in electron density of the ionized interstellar medium (ISM) along the kiloparsec distances to pulsars from the local fluctuations in electron density due to variations in the solar wind.

The NANOGrav 11-year Data Set consists of high-precision time-of-arrival (TOA) measurements 45 millisecond pulsars spanning up to 11.4 years.  Each pulsar was observed at approximately a monthly cadence, over widely separated radio frequencies in order to measure both TOAs and dispersion of the pulsar signal at each observing epoch.  Observations were made using the Arecibo Observatory and the Green Bank Observatory.  Typical TOA measurement precision was in the range $0.1$ to 1~$\mu$s. Further details of the data set are in \citetalias{abb+18a}.

\subsection{Monitoring of Solar Electron Density}
Since the initial use of astrophysical radio sources to measure solar wind density via scintillation \citep{hewish+1967}, a program for monitoring solar wind densities, especially at higher solar latitudes, has continued  \citep{coles1978,manoharan2010,tokumaru+2013}. In addition to ground-based monitoring, a considerable number of resources have been used to study the solar wind from space. Many of these spacecraft have included instruments for electron density measurements, including Ulysses's Solar Wind Observations Over the Poles of the Sun \citep[SWOOPS;][]{ulysses92}, the Orbiter Retarding Potential Analyzer (ORPA) instrument mounted on the NASA Pioneer Venus Orbiter spacecraft \citep{pioneer_orpa1980} and the Advanced Composition Explorer (ACE) with its Solar Wind Electron Proton Alpha Monitor \citep{aceswepam1998}. Similar to pulsar timing measurements, the Viking mission used dual frequency delays to make early measurements of the latitudinal dependence of the solar electron density \citep{ma81}. The Parker Solar Probe \citep{bale+2016} with its Solar Wind Electrons Alphas and Protons (SWEAP) instrument \citep{kasper+2016} is revolutionizing our understanding of the solar wind in the inner solar system \citep{bale+2019}.

Apart from the Viking data, these space missions take {\it in situ} measurements of the solar electron density. These allow for one variety of long-term monitoring of the solar wind and Ulysses data have been used to study its structure and behavior at high to mid latitudes \citep{issautier+01}. Obviously, this monitoring requires the use of costly spacecraft, and while the data from pulsar timing does not provide the same type of fine-grained local spatial information, the many lines of sight to pulsars allow for omnidirectional integral monitoring of the solar electron density from ground-based facilities and the scrutinization of various models for its structure and evolution. PTAs provide a probe of the solar wind that is distinct from and complementary to space missions.

\subsection{Noise Mitigation in PTAs}

The main goal of PTA experiments is to use precise long-term measurements of millisecond pulsars to observe gravitational-waves (GWs) in the nanohertz regime. Many pulsars are used since the unique noise properties of individual pulsars necessitates the corroboration of common\footnote{``Common" is used here in the sense that the signal is present, in whole or in part, in observations of all pulsars in the timing array.} astrophysical signals across multiple sources and because lower-amplitude common signals can be drawn out of the noise of more pulsars. Additionally, PTAs allow us to search for other common signals, for instance the motion of the solar system barycenter \citep{vts+20} or errors in terrestrial time standards \citep{IPTAtime20}. 

One motivation for this work is to develop a pan-PTA solar wind model as one component of the next generation of PTA noise models. Dispersion measure (DM) variations are one of the largest noise sources in PTA data, \citep{lcc+2016,css16,cs10,jml+2017}, and modeling these variations is an important part of PTA noise mitigation strategies---it is, in fact, a major driver of PTA observing strategies \citep{nanograv_astro2020, lmc+2018}. DM values in \citetalias{abb+18a} vary from $\sim3-300 \;{\rm pc}/{\rm cm}^3$, while the variations are usually on the order of $10^{-3} \;{\rm pc}/{\rm cm}^3$. While a piecewise binned model of DM variations \citep[DMX, see][]{abb+15} has served NANOGrav well, it removes up to a third of the power through the timing model transmission function \citep{hazboun:2019sc,abb+15,jml+2017}, a problem endemic to any DM variation model that is part of the timing model fit. Additionally it has been shown that asynchronous multi-band observations can lead to misestimation of DM variations \citep{lcc+15} when binned together, either in DMX or an interpolation basis. \citet{niu+2017} demonstrated that observations only a day apart can be insufficient to remove sharp solar wind effects, or ``cusps". In \citet{hazboun:2020slice} and \citet{leg+2018} the effect of sharp unmodeled cusps in DM variations was demonstrated. Mismodeling of solar wind cusps could then also present as broadband white noise in pulsar datasets. In particular, this could adversely affect GW searches for single sources which are especially dependent on the high frequency noise floor of PTAs \citep{lam18_optcad,lam+18optfreq,lam+21_iar}.

Perhaps most importantly for PTAs, in addition to the noise introduced by mismodeling DM variations, it was shown in \citet{thk+2016} that the solar wind can actually manifest spatial correlations amongst the pulsars, a potential source of confusion when searching for GWs, especially a GW background which presents in the data as a low-frequency spatially-correlated process. 
In this paper, we demonstrate how a fully Bayesian framework, built upon the extensive analysis infrastructure developed by PTA collaborations to search for GWs, can isolate the signal from the solar wind in PTA data. This allows for precise noise mitigation as well as monitoring of the overall solar electron density both as a function of time and observational LOS.

The paper is organized as follows. In \mysec{sec:solar_electron} the DM parameter will be derived along with various analytical models for a spherically symmetric and static solar electron density. In \mysec{sec:bayes_model} we present the full Bayesian model for pulsar timing data and how we incorporate our solar wind models. In \mysec{sec:sw_models} three models for the behavior and structure of the solar electron density will be presented along with the results of using these models to analyze \citetalias{abb+18a}. In \mysec{sec:dis} we discuss the incorporation of these models into PTA data analyses as well as how these models might be used in the future to bolster efforts to study solar physics.

\section{Solar Electron Density Models} \label{sec:solar_electron}
The simplest and most common model for the solar electron density used in pulsar timing software packages like \tempo, \tempotwo and \pint \citep{Nice:2015a, tempo2, luo+2021} is the spherically-symmetric, time-independent expression $n_{e}(\vec{r})=n_E\left(1\, {\rm au}/r\right)^2$, where $n_E$ is the electron density at $1\, {\rm au}$, herein measured using units of $1/{\rm cm}^3$ to match pulsar timing software. 
\begin{figure*}
    \centering
        \includegraphics[width=0.45\textwidth]{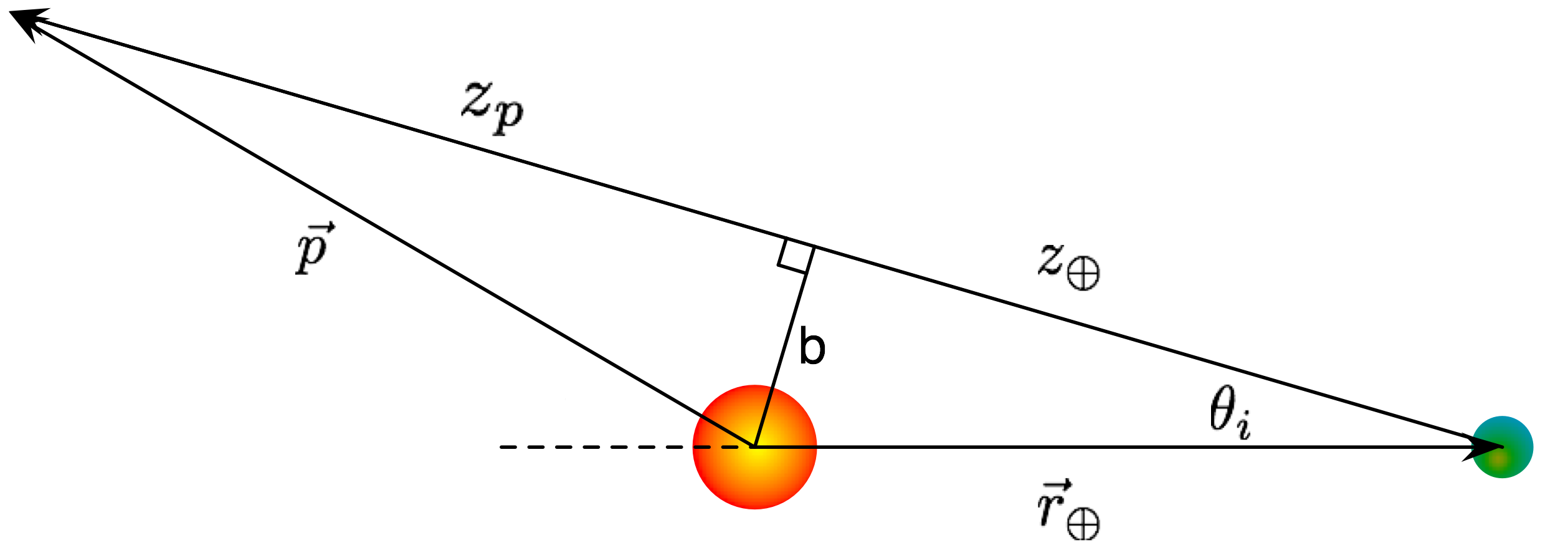}
        \caption{Solar Wind Geometry: The figure defines the variables used for the solar wind electron density line integral. Here we have adopted some of the nomenclature of scattering calculations, as in \citet{aksim+2019}, using $b$ as the impact parameter and defining $\theta_i$ as the ``impact angle'', i.e the solar elongation. The vector $\vec{p}$ points from the Sun to the pulsar, while $\vec{r}_\oplus$ points from the Sun to the observatory on Earth. The distance to the pulsar is effectively infinite when compared to the distance to the Earth, so $z_p$ and  $\vec{p}$ are effectively parallel. \label{fig:sw_geometry}}
\end{figure*} 
The DM parameter used in pulsar timing is the column density, i.e. the integral of $n_e(\vec{r})$ along the line-of-sight to the radio source
\begin{eqnarray}
{\rm DM}_\odot&=&\int_{r_\oplus}^{r_{p}}n_e\left(\vec{r}\right)ds\\
		       &=&\int_{r_\oplus}^{r_{p}}n_E\left(\frac{1\,{\rm au}}{r}\right)^{2}ds\\
                        &=&n_E\left(1\,{\rm au}\right)^{2}\frac{1}{b}\left[\frac{\pi}{2}+\tan^{-1}\left(\frac{z_{\oplus}}{b}\right)\right]\\
                        &=&n_E\left(1\,{\rm au}\right)^{2}\frac{\pi-\theta_{i}}{r_\oplus\sin\theta_{i}}\label{eq:dm_r2},
\end{eqnarray}
where the first line is the generic expression, and the last equality uses the relationship $\frac{z_{\oplus}}{b}=\cot(\theta_i)=\tan\left(\frac{\pi}{2}-\theta_{i}\right)$ and $b=r_\oplus\sin\theta_{i}$. See \myfig{fig:sw_geometry} for the definitions of these variables\footnote{Note that the $\rho$ used in \citet[(and so forth)]{edwards+06,tiburzi+2019} and $\theta_i$ are supplementary angles, i.e., $\rho=\pi-\theta_i$.}. The time delay is then dependent on the radio frequency and the DM by
\begin{equation}
\Delta t=\frac{e^2}{2\pi m_e c^2}\frac{\rm DM}{\nu^2}\,.
\end{equation}
DM is usually measured in units of ${\rm pc}/{\rm cm}^3$, and the constants of nature in front combine to $4.15\times10^3 \,{\rm MHz}^2 \,{\rm pc}^{-1} \,{\rm cm}^{3}  \,{\rm s}$.

In \citet{you+2007} the authors use a two phase model assuming that the LOS to the pulsar crosses the fast and slow solar wind. Higher order terms, in addition to the simple $1/r^2$ model are used to ameliorate the effect of realistic solar wind time delays. The slow solar wind model\footnote{Here we have converted the expressions in \citet{you+2007} to units more familiar to pulsar timing astronomers.},
\begin{eqnarray}\label{eq:ne_slow}
n_{\odot}\times {\rm cm}^{3}&=&8.867\left(\frac{1\,{\rm au}}{r}\right)^{2}+1.186\left(\frac{1\,{\rm au}}{r}\right)^{2.7}\nonumber\\
                                               &&+1.517\times10^{-6}\left(\frac{1\,{\rm au}}{r}\right)^{6}\nonumber\\
                                               &&+1.431\times10^{-29}\left(\frac{1\,{\rm au}}{r}\right)^{16}
\end{eqnarray}
was developed in \citet{you+2007} by combining the near and far distance models in \citet{allen1947} and \citet{ma81}. A separate model was used for the fast solar wind 
\begin{eqnarray}\label{eq:ne_fast}
n_{\odot}\times {\rm cm}^{3}&=&2.498\left(\frac{1\,{\rm au}}{r}\right)^{2}\nonumber\\
					     &&+1.860\times10^{-4}\left(\frac{1\,{\rm au}}{r}\right)^{4.39}\nonumber\\
                                               &&+4.067\times10^{-30}\left(\frac{1\,{\rm au}}{r}\right)^{16.25}
\end{eqnarray}
developed in \citet{GuhathakurtaFisher1998,GuhathakurtaFisher1995}. These models were then combined with measurements of the portion of the line-of-sight in the fast and slow wind to better estimate the cusps to the pulsar timing data. In \citet{you+2012} the fraction parameter for the two phases of the model is fit for as part of the analysis. Additionally a time dependent component is investigated. The model in \citet{you+2007,you+2012} has potential as a useful tool for removing the solar wind signal from pulsar timing data, especially for single pulsars, where disentangling the DM variations from the ISM is difficult. 

Recently, however, \citet{tiburzi+2019,tiburzi+2021} showed that the simple $1/r^2$ model for $n_E$ commonly used in pulsar timing software packages and the \citet{you+2012} model are both insufficient for accurately modeling dispersion delays from the solar wind. Even with allowances for time variability, these models still fall short of the mitigation needed for the low frequency radio data taken by the LOw-Frequency ARray (LOFAR) \citep{lofar2011,lofar2013} used in the analyses. 

In \citet{aksim+2019} the authors compare their measurements of DM using Very Long Baseline Interferometry (VLBI) observations of quasars to both a spherically symmetric model and a numerical integral of the Alfv\'en Wave Solar Model (AWSoM) \citep{awsom2014}. As in other studies the AWSoM matches well with the astrophysical measurements, however this model is currently too cumbersome to be used effectively by PTAs. \citet{aksim+2019} also fit for a variable power spherically symmetric model and find reasonable agreement with their time delay measurements. 

\subsection{General Spherically Symmetric Solar Wind Density}\label{subsec:gen_solar}
Here we derive generic expressions for the dispersion measure for any spherically symmetric electron density fall off. While these expressions will assume the source of the electrons is the sun, they are general enough to be used for any spherically symmetric source of streaming electrons along the LOS to a pulsar, a stellar wind from a star orbiting a pulsar, for instance. Relaxing the assumption that the electron density around the sun drops off as $1/r^{2}$, we model the drop off with a more generic power law dependence. These calculations are similar to those in \citet{aksim+2019}, where the analytical expression for the time delay was calculated directly. 

Assuming that $p$ is the exponent in the power law dependence we go through a similar calculation as in \myeq{eq:dm_r2},
\begin{eqnarray}
\text{DM}_{\odot}^{p}	&=&	\int_{\text{Earth}}^{\text{Pulsar}}n\left(\vec{r}\right)ds=\int_{r_\oplus}^{r_{p}}n^{(p)}_{E}\left(\frac{1\,{\rm au}}{r}\right)^{p}ds\\
         &=&\int_{-z_\oplus}^{z_{p}}n^{(p)}_{E}\left(\frac{1\,{\rm au}}{\sqrt{b^{2}+z^{2}}}\right)^{p}dz\\
	&=&	n^{(p)}_{E}\left(\frac{1\,{\rm au}}{b}\right)^{p}\left[z\times\,_{2}\mathcal{F}_{1}\!\left(\frac{1}{2},\frac{p}{2},\frac{3}{2},\text{-}\frac{z^{2}}{b^{2}}\right)\right]_{\text{-}z_\oplus}^{z_{p}}
\end{eqnarray}
where $_{2}\mathcal{F}_{1}$ is a hypergeometric function. This expression can be further simplified by assuming the pulsar is sufficiently distant that we can take the limit $z_{p}\rightarrow\infty$. This gives
\begin{equation}\label{eq:dm_rp}
\text{DM}_{\odot}^{p}	=	n^{(p)}_{E}\left(\frac{1\,{\rm au}}{b}\right)^{p}\left(z_\oplus\times\,{}_{2}\mathcal{F}_{1}\!\left(\frac{1}{2},\frac{p}{2},\frac{3}{2},\text{-}\frac{z_\oplus^{2}}{b^{2}}\right)+\frac{b\sqrt{\pi}}{2}\frac{\Gamma\left(\frac{p}{2}-\frac{1}{2}\right)}{\Gamma\left(\frac{p}{2}\right)}\right)\; ,
\end{equation}
where $\Gamma\left(x\right)$ is the Gamma function. Compare this expression to the time delay equation given in \citet{aksim+2019}. \myeq{eq:dm_rp} simplifies to \myeq{eq:dm_r2} when $p=2$. See Appendix~\ref{sec:highpowers} for expressions with higher integer values of $p$. The relationships below \myeq{eq:dm_r2} can be used to write this in terms of $\theta_i$. One can choose to build a solar electron density model using multiple summed terms, as in \citet{you+2007,you+2012,allen1947,ma81,GuhathakurtaFisher1998,GuhathakurtaFisher1995}, or allow the index to vary, as in \citet{aksim+2019}.

\section{Bayesian Methods} \label{sec:bayes_model}

The solar wind modeling presented here relies on the analysis infrastructure developed by PTAs for GW searches \citep{vanHaasteren:2012hj,Lentati:2016ygu,taylor2013,vHv:2014,vanhaasteren2009}. 
While a gravitational wave background only causes residuals of tens of nanoseconds, the solar wind signal is much stronger ($\sim1.7 \mu {\rm s}$ at 1 GHz with $\theta_i=10^\circ$). This allows us to undertake an in depth study of the solar wind by modeling the solar electron density across all of the pulsars simultaneously. 

To decouple the variations within the ISM along the LOS to each of the pulsars from the solar electron density that is varying locally, we use a deterministic solar wind model based on the expressions in \mysec{subsec:gen_solar}. The solar wind electron density is fit as a global parameter across all pulsars, while the geometry of the solar wind model is dependent on the details of the individual Sun-Earth-Pulsar angles, $\theta_i$, shown in \myfig{fig:sw_geometry}. In addition to this model each pulsar is also fit with a quadratic polynomial in time over the data set for the DM, encoded with the pulsar timing parameters, \texttt{DM1} and \texttt{DM2}, as well as an additional Gaussian process model that emulates the variations of the ISM, which will be discussed in \mysec{subsec:dm_vars}.

The PTA analysis framework includes a timing model marginalization, based on a linearized timing model \citep{vanHaasteren:2012hj} and a full noise treatment that includes a generalized covariance matrix, parametrized by TOA errors, three additional white noise parameters and a power law red noise model \citep{Lentati:2016ygu,taylor2013}. The PTA likelihood model is implemented using the \enterprise software suite  \citep{enterprise} and the various models and extensions compiled in \extensions \citep{e_e}. We use the Parallel Tempering Markov-Chain Monte Carlo Sampler, \texttt{PTMCMCSampler}, for numerically integrating our likelihood \citep{ptmcmc}. 

It should be noted that while this formalism is focused on high precision millisecond pulsars, the code infrastructure can be used on data from any type of pulsar and its timing data. Observations of the more numerous canonical pulsars may be useful for these types of solar investigations in the future.

\subsection{ACE SWEPAM Prior}\label{subsubsec:ACEprior}
\begin{figure}
    \centering
        \includegraphics[width=0.45\textwidth]{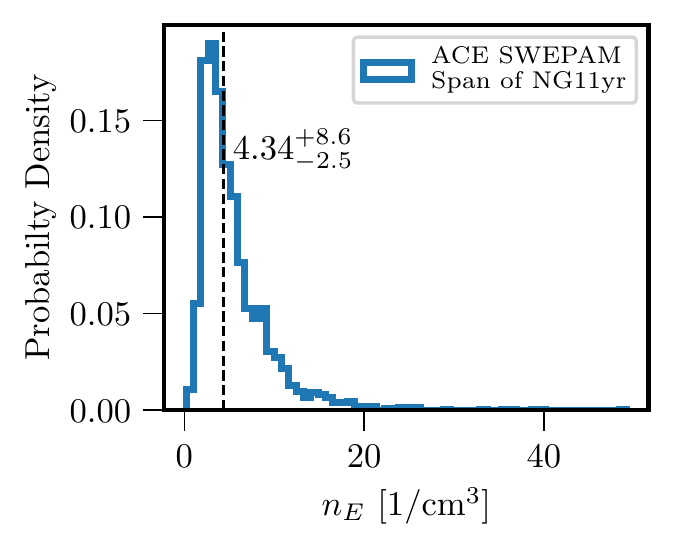}
        \caption{ACE SWEPAM Prior. We use in situ measurements of the solar electron density at Earth's orbit to construct a prior for our analysis. Note that the prior extends all the way up to $\sim50{\rm \,cm}^{-3}$. \label{fig:ace_prior}}
\end{figure}

As mentioned in \mysec{sec:intro}, a number of past and ongoing space missions have collected solar electron density data over the past few decades. In order to take advantage of this large base of knowledge about the solar electron density we used the solar electron density data from the ACE Solar Wind Electron Proton Alpha Monitor \citep{aceswepam1998} to build an informative prior for $n_E$, the solar electron density at $1\,{\rm au}$. We binned the data from the same time span as \citetalias{abb+18a}, corrected the density from the L1 Lagrange point using a $1/r^2$ density model\footnote{$n_e^{\rm ACE}=n_E\left(\frac{1\;{\rm au}}{0.997\,{\rm au}}\right)^2$} and used an empirical distribution to build a random variable in \texttt{SciPy}. The distribution and median are shown in \myfig{fig:ace_prior}. 
The ACE SWEPAM {\it in situ} data are a useful prior for limiting the extent of parameter space at which we expect to be working, roughly $0.01-50 {\rm \,cm}^{-3}$. However, the electron flow in the ecliptic plane is known to be different than the fast flow at other solar latitudes \citep{ma81,ulysses92,GuhathakurtaFisher1998}. We do not expect the ACE measurements to match our results perfectly, but it is a good comparison in these analyses as a sanity check. In practice the prior was only informative when compared to the data in some individual pulsar runs \textemdash\ usually pulsars without strong solar wind signals in their TOAs. In tests, the full PTA runs retrieved the same posteriors from uninformative uniform priors, but we use the  ACE prior throughout the work presented here to decrease convergence times in our Markov Chain Monte Carlo runs.

Uniform priors were used for the spectral index in all higher order searches and log-uniform priors were used for $n^{(p)}_E$ priors when $p\neq2$.

\subsection{Dispersion Measure Variation Model}\label{subsec:dm_vars}

Gaussian processes (GPs) have been used extensively to model time-correlated noise in pulsar datasets \citep{abb+2020gwb,abb+18b,goncharov+21a}. Commonly, they are used to model the GW background, achromatic red noise, such as that caused by intrinsic spin noise, and chromatic noise due to the bulk movement of the ISM \citep{goncharov+21a,Lentati:2016ygu}. 

The standard prior function of GPs in these datasets is a red (negative spectral-indexed) power law model with a Fourier-basis. The solar wind has strong spectral characteristics at Fourier frequencies of $1/{\rm yr}$ and higher modes \citep{madison+2019}. In early testing these modes showed themselves to be highly covariant with some frequencies in a Fourier-basis GP, so we have instead adopted for our DM variations model another common ansatz for GPs: a square exponential kernel,
\begin{eqnarray}
    k_\mathrm{SE}(t_1,t_2) 
    =& \sigma^2  \exp{\left(-\frac{| t_1-t_2|^2}{2\ell^2} \right)} + \left(\frac{\sigma}{500}\right)^2,
\end{eqnarray}
where $\sigma$ is an overall variance, $\ell$ is a time scale of variation, and a constant variance is also added as a regularizing term for stable inversions and to ensure a minimal value of the variance. This kernel represents the realization average of covariance matrices described by this type of variation.  They are similar to the structure functions used in the pulsar timing literature for characterizing noise due to variations in the ISM---for instance, see Equation~(7.1) in \citet{fc90} or Section~5 of \citet{lcc+2016}. Such structure functions were also used as a check in \citet{tiburzi+2019} to ensure the iterative process used therein full disentangled the solar wind signal form the DM variations in the ISM. Our Bayesian framework allows for simultaneous fitting across multiple pulsars and frees us from the iterative process used in this latter publication. \citet{tiburzi+2021} also uses a Bayesian analysis to separate these signals, but only within single pulsar datasets in order to keep track of solar wind dependencies on ecliptic latitude. The GP model above (along with our solar wind model) replaces the use of a short-time-span DMX model normally used in NANOGrav datasets \citep{abb+18a,aab+20nb,aab+20wb}, since DMX cannot distinguish between the ISM and solar wind contributions to the measured DM value at each epoch. Here a standard $15$ day linear interpolation basis in the time domain is used for constructing realizations of the DM variations. 

\subsection{Noise Analyses}\label{subsec:noise}
Since we are implementing entirely new DM variation models from the original \citetalias{abb+18a} analysis, we redid single pulsar noise analyses using our solar wind and DM GP models. This allowed us to recover new values for the standard white noise parameters, EFAC, EQUAD and ECORR, \citep[see e.g.,][]{abb+15,lcc+2017} and power law achromatic red noise parameters for each pulsar. These single pulsar noise runs then consist of a set of three white noise parameters for each backend/receiver combination used for observations, two DM GP parameters, $\sigma$ and $\ell$, a solar electron density, $n_E$, and two power law red noise parameters, $A_{\rm RN}$ and $\gamma$. 

Using only one pulsar's dataset it is impossible to fully disentangle the solar wind electron density from the ionized ISM. Any given observation's DM variation can be attributed to either changes in the interplanetary {\it or} interstellar media. The key to these analyses is a global fit where the DM GP and red noise parameters are allowed to vary individually for each pulsar, while the solar electron density is set as a global parameter across all of the pulsars. Since pulsars have upwards of $20+$ white noise parameters it is cumbersome to vary all of these parameters across the entire PTA, so we adopt the usual practice in PTA analyses \citep{abb+2020gwb} of setting their values constant in the full PTA analysis following single-pulsar noise modeling.

\section{Bayesian Model Implementation and Analysis}\label{sec:sw_models}

We use the expressions in \mysec{subsec:gen_solar} to build models for a spherically symmetric solar electron density and analyze \citetalias{abb+18a}. We explore three different models for the behavior and structure of the solar electron density: higher order spherically symmetric solar wind models, time-binned solar wind density models, and continuous time-dependent solar wind perturbations. Each model demonstrates how our full Bayesian framework provides access to the unique probes of the solar wind available in PTA datasets. 

\subsection{Higher Order Models}
Any number of higher order spherically symmetric models can be constructed using the tools developed here. We present only a few to demonstrate the feasibility of detecting these higher order signals using PTA datasets. In general we follow the methods of \citet{aksim+2019} and fit for a varying power dependence, $\sim1/r^p$, of the solar electron density, though we still investigate models of the type in \citet{you+2007,you+2012} with multiple summed fixed-power terms. The fit for the proportion of the two-phases (slow and fast solar wind) in \citet{you+2012} is dependent on the individual lines-of-sight for the individual pulsars which is out of the scope of the current tests for these modeling tools. 
\renewcommand{\arraystretch}{1.3}
\begin{deluxetable*}{c|cc|cc|cc}
\tabletypesize{\small}
\tablewidth{0.9\textwidth}
\tablecolumns{7}
\tablecaption{Table of higher order spherically symmetric solar wind density terms. Parameters without sub(super)scripts are set constant. The median of free parameters is reported along with the $68\%$ credible interval. Parameters with a superscript $^{95\%}$ represent $95\%$ upper limits of the parameter.\label{tab:higher_order}}
\tablehead{\colhead{} & \multicolumn{2}{c}{First Term}  & \multicolumn{2}{c}{Second Term} & \multicolumn{2}{c}{Third Term} }
\startdata
Model & \phs$p$ & $n^{(p)}_E\;[{\rm cm}^{-3}]$ & \phs$p$ & $\log_{10}n^{(p)}_E$  & \phs$p$ & $\log_{10}n^{(p)}_E$ \\
\hline
1	&\phs$2$					&$6.9^{+0.13}_{-0.13}$	&\multicolumn{2}{c|}{---}								&\multicolumn{2}{c}{---}\\
2	&\phs$2.29^{+0.01}_{-0.02}$	&$4.7^{+0.20}_{-0.20}$	&\multicolumn{2}{c|}{---}								&\multicolumn{2}{c}{---}\\
3	&\phs$2$					&$2.5^{+0.56}_{-0.63}$	&\phs$2.4^{+0.04}_{-0.05}$ 	&\phs$0.38^{+0.10}_{-0.07}$	&\multicolumn{2}{c}{---}\\
4	&\phs$2.30^{+0.02}_{-0.01}$	&$4.5^{+0.18}_{-0.22}$	&\phs$4.3^{+0.58}_{-0.73}$ 	&\phs$-4.1^{95\%}$			&\multicolumn{2}{c}{---}\\
5	&\phs$2$					&$6.8^{+0.18}_{-0.19}$	&\phs$4.39$				&\phs$-2.87^{+0.03}_{-0.02}$	&\phs$16.25$	& \phs$-24.8^{95\%}$
\enddata
\end{deluxetable*}

In \mytab{tab:higher_order} we summarize results from a number of similar analyses. The aim here was not to exhaustively test the various powers from \myeq{eq:ne_slow} and \myeq{eq:ne_fast}, as the real model is expected to be a mixture of the two \citep{you+2007}, but rather to test the sensitivity of \citetalias{abb+18a} to some of these higher order terms. \mytab{tab:higher_order} shows a progression from the overly simplistic $1/r^2$ model, where we find $n^{(2)}_E=6.9^{+0.13}_{-0.13}\frac{1}{{\rm cm^3}}$, to models with more terms, and more free parameters. Model 2, where the exponent in the density relation is allowed to vary, shows that there is support for a higher order model for the solar electron density, $p=2.29^{+0.01}_{-0.02}$. This power is in agreement with the best fit power from \citet{aksim+2019} of $p=2.3$. Their value of $n^{2.3}_E=2.59\pm0.13\frac{1}{{\rm cm^3}}$ is smaller than our recovered value, but the \citet{aksim+2019} value is from only one observation, while our value is an $11.4$ year average. We will see in \mysec{subsec:sw_bins} that time varying values can differ by up to a factor of 4 or more. 

Models 3 and 4 show that there is broad support for a second term in the density model when the power of the first term is kept constant, but the analysis only retrieves upper limits for a third term when the first two are kept constant. These analyses broadly support what has been previously shown in \citet{you+2012,you+2007} that pulsar timing data, especially for an array as we are treating here, is sensitive to a much more complicated model of the solar electron density than is routinely assumed. Model 5 shows the limits of the data set's sensitivity, as we do not seem to detect a third term with the large index predicted in \citet{you+2007}.

\subsection{Time-Binned Solar Wind Density}\label{subsec:sw_bins}

\begin{figure*}
    \centering
        \includegraphics[width=0.95\textwidth]{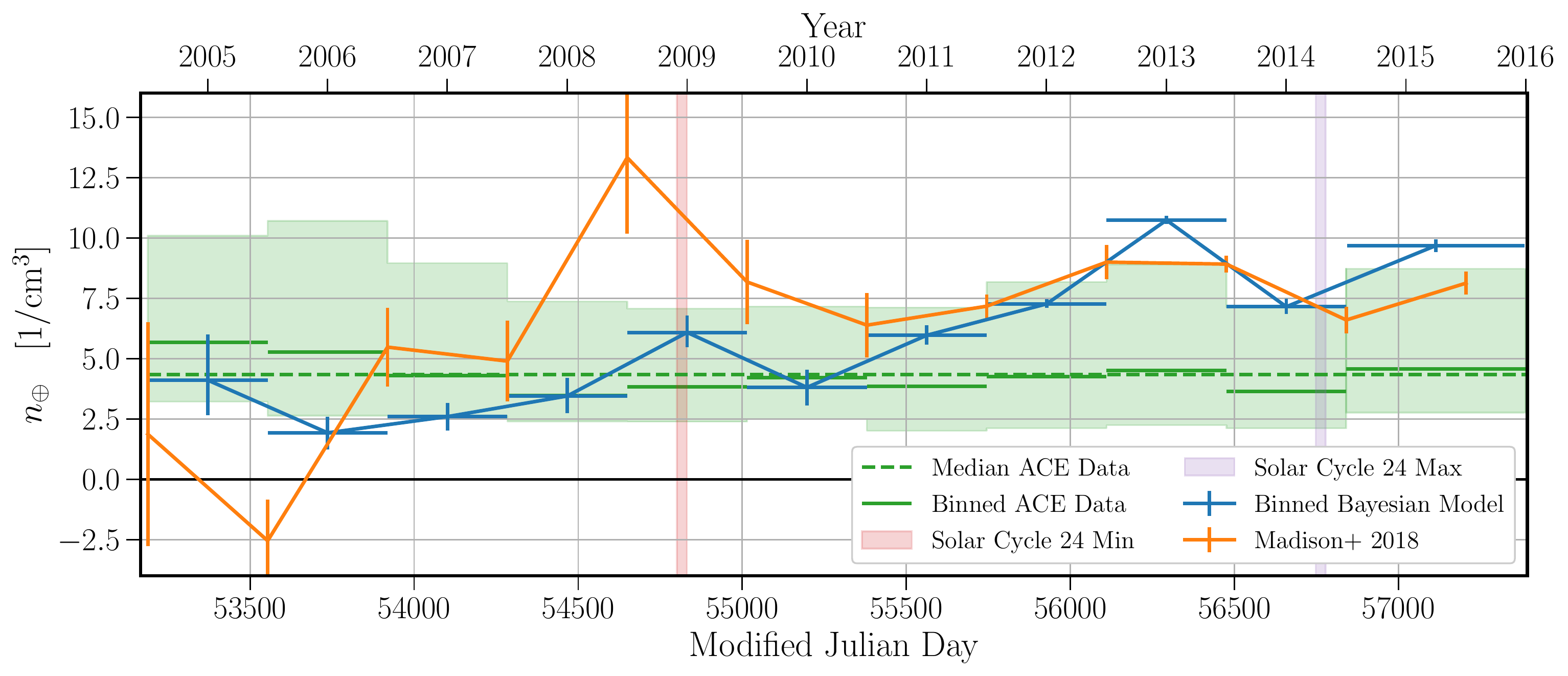}
        \caption{Year binned solar electron density. The blue plusses show the values of our binned solar electron density search. The horizontal error bars show the width of the bins used, mostly one year here, except the last bin which is 1.4 years long. The orange lines with vertical error bars show the results from \citet{madison+2019} that used this same dataset, \citepalias{abb+18a}. The green shaded blocks show median and 68\% confidence intervals from the {\it in situ} measurements of the ACE satellite's SWEPAM instrument. The pink and purple shaded vertical regions show the minimum and maximum, respectively, months of solar cycle 24.\label{fig:madison}}
\end{figure*}

As in \citet{madison+2019} and \citet{tiburzi+2021}, we search for separate values of $n_E$ with $p=2$ during different periods of time by fitting values of $n_E$ for discrete time bins of the dataset. In the \citet{madison+2019} analysis the bins are set to be a year long and a fit is done to the DMX times series of the pulsars. DMX is a piece-wise time-binned analysis of the DM variations, \citep[see, e.g.][]{abb+15}. We present our results and those of \citet{madison+2019} in \myfig{fig:madison}, along with binned ACE SWEPAM data for the same time period. The ACE data are {\it in situ} measurements of the local electron density, and so can be considerably noisier than the large effective average taken by the column density measurements to which the pulsar timing data are sensitive. In addition, since ACE orbits in the ecliptic plane, the SWEPAM instrument is really only sampling the slow solar wind, and would not track changes in the higher altitude fast wind. This is supported by scintillation studies that track high altitude solar wind density, see for instance \citet{porowski+2021} for a model based on scintillation data. Their model for solar electron density shows strong positive correlations on the high latitude wind density with the solar cycle. Lastly, the ACE data are taken by the same instrument across the 11+ year time span shown here, while the NANOGrav PTA continually increased the number of pulsar in the array (hence LOSs) and sensitivity, including a ``high cadence'' observing campaign in the latter portion of the dataset \citep{abb+18a}.

The largest differences between our values and those of \citet{madison+2019} occur in the first part of the dataset where the observation cadence was significantly less.
The fit in \citet{madison+2019} was done using a $\chi^2$-minimization, and did not include any priors for the solar electron density. Hence the unphysical negative value of the second bin is a result of the $\chi^2$-minimization. The large value in the fifth bin occurs during a time when there is a gap in the PTA data. This will be explored more in the finer binning used next. The \citet{madison+2019} analysis is not as robust as the analysis here, since rather than fitting a secondary data product (DMX) we are fitting the solar wind as part of a full PTA-wide analysis, which includes a noise model fit and timing model marginalization.

\begin{figure*}
    \centering
        \includegraphics[width=0.97\textwidth]{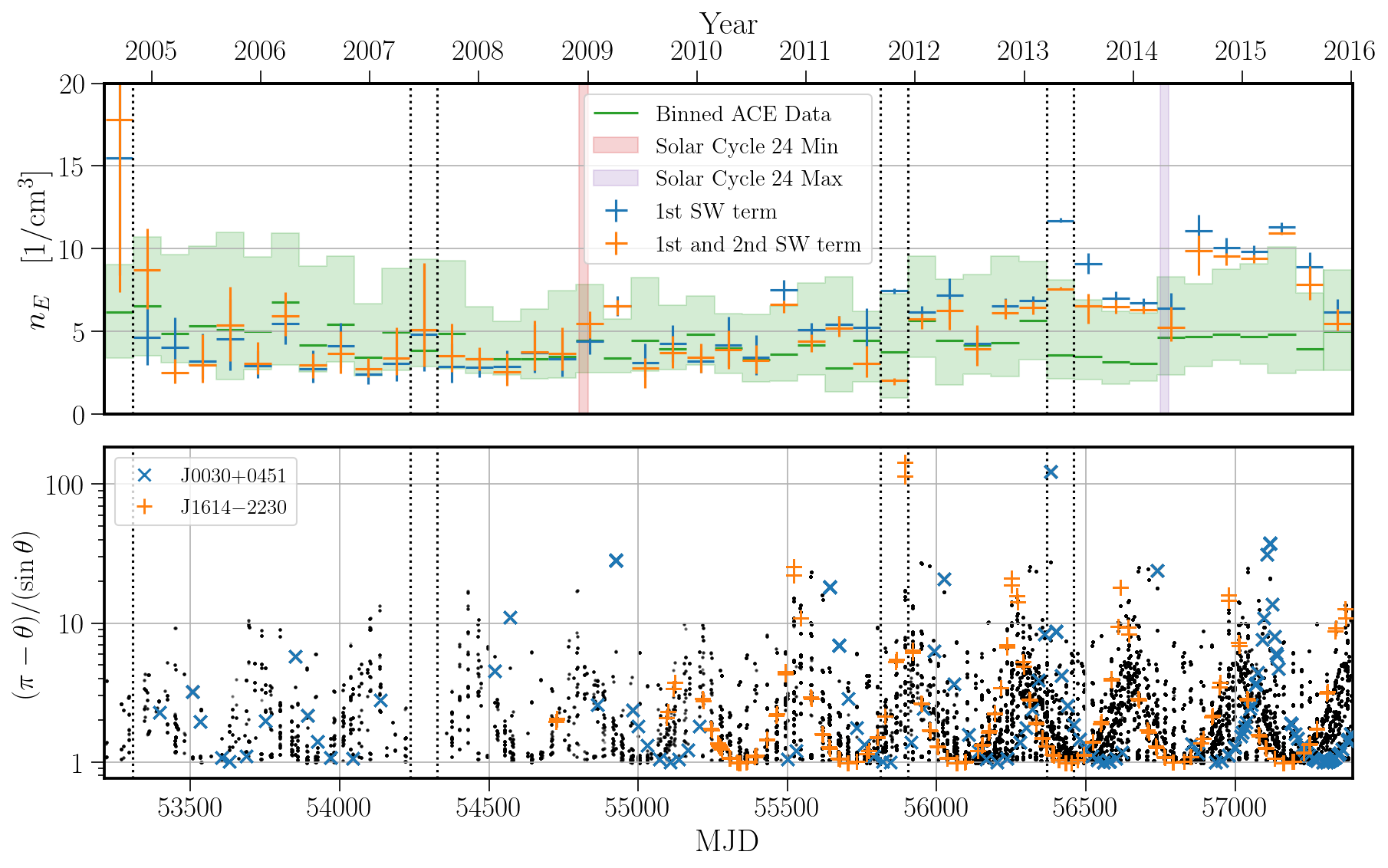}
        \caption{Three month binned solar electron density. The horizontal error bars show the width of the bins used, while the vertical error bars show the inner $68\%$ credible region of the binned $n_E$ posteriors. The pink and purple shaded vertical regions show the minimum and maximum, respectively, months of solar cycle 24. The vertical dotted lines show the boundaries of a few important bins. The first shows the end of the first bin which contains the fewest number of TOAs. The bin near MJD $54250$ has no TOAs due to down time at both the Arecibo Observatory and the Green Bank Observatory. The dotted vertical lines near  MJD $55800$ and MJD $56400$ show bins where the values for $n_E$ in the two models differ by a substantial amount. The bottom panel shows the geometric part of the solar DM obtained from a $1/r^2$ model for the solar wind density calculated for all TOAs in the dataset. Note the log scale on the y-axis. Only two pulsars, J0030+0451 and J1614$-$2230, have geometric factors larger than $100$ and the factors for these have been highlighted with colored x's and +'s, respectively. \label{fig:3month_bin}}
\end{figure*}

In \myfig{fig:3month_bin} we show Bayesian binned results for bin sizes of three months again with $p=2$, showing that \citetalias{abb+18a} has enough information to fit $n_E$ with finer resolution than is done in \citet{madison+2019}. The $68\%$ credible intervals for the binned values are small compared to the same intervals in the ACE data. In addition to the binned fit for $n_E$, we also run an analysis with a second density, $n^{(4.39)}_E$, in the solar wind density profile model with $p=4.39$.  We chose this index from the fast solar wind model of \myeq{eq:ne_fast} in order to investigate how a higher order term might effect the binned values of $n_E$ specifically from the few pulsars with very small $\theta_i$. We use a time-constant value of $n^{(4.39)}_E$ in this application for ease of analysis. 

This second analysis, with an $n^{(4.39)}_E$ parameter highlights the importance of these higher order terms in the density model when there are observations with lines of sight close to the Sun. The geometric factor for the ${\rm DM}_\odot$ delay, $(\pi-\theta_i)/\sin \theta_i$, of all TOAs is shown in the bottom panel of \myfig{fig:3month_bin}. There are two occasions in the 11-year dataset where $\theta_i$ is very small for PSR J0030+0451 and PSR J1614$-$2230. These can be seen as the points where the geometric factor is larger than $\sim100$. The two bins, marked by vertical dots, where these occur correspond to the two bins where the recovered values for $n_E$ differ by many standard deviations from the usual $1/r^2$ model for the density, over estimating $n_E$. While the $p=4.39$ model seems to ameliorate the effect of these small $\theta_i$ observations, the direct cause of the anomalously high measurements of $n_E$ could also be due to the highly inhomogeneous streaming of the solar wind at these small distances or from individual events, i.e. coronal mass ejections. See Appendix~\ref{sec:cme} for more discussion.

Two other bins warrant attention. The first bin has only a few TOAs with narrow frequency coverage, and hence parameters recovered have large error bars. The bin near MJD 54250 has no TOAs due to concurrent down time at both the Arecibo Observatory and the Green Bank Observatory. In this bin, the ACE prior is returned in the posterior. Hence no information beyond the prior is gained.

\subsection{Continuous Time-Dependent Solar-Wind Perturbations}\label{subsec:sw_gp}
In addition to the piecewise constant models for $n_E$ discussed above, we also implemented a continuous Fourier-basis model for $n_E(t)$ as a perturbation to a mean $n_E$. The model is reminiscent of the free spectral models used by PTAs to describe achromatic red noise \citep{abb+2020gwb}. It is constructed using the TOAs for each pulsar, parametrized with a separate value for each component of the basis,
\begin{equation}\label{eqn:n_fourier}
n^{F}_e(t) = \sum^{N_f}_{j=0}a_j\sin(2\pi f_j t)+b_j\cos(2\pi f_j t)\; .
\end{equation}
Rather than trying to reproduce the fine structure recoverable by the \citet{you+2012} model, here we are concerned with long to mid scale variations of $n_E$, therefore we use a set of frequencies based on the time span, $T=11.4$~yr, of \citetalias{abb+18a} and include 30 linearly spaced frequencies ranging from $[1/T,30/T]$, using 2 parameters per frequency, $a_j$ and $b_j$. 

\begin{figure*}
    \centering
        \includegraphics[width=0.97\textwidth]{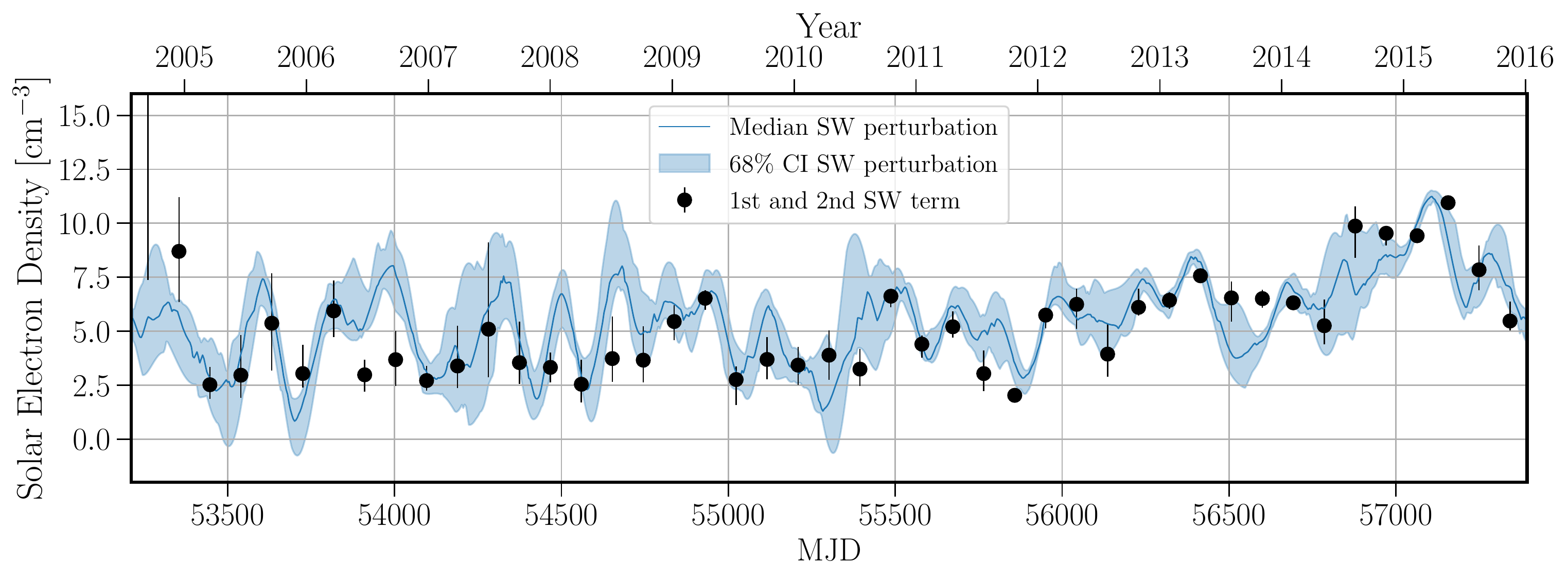}
        \caption{Continuous solar wind density model. The black dots show the same binned values for $n_E$ shown in orange in the top panel of \myfig{fig:3month_bin}. The blue trace and shading show the median and inner $68\%$ credible interval of 1000 realizations of the continuous solar electron density perturbation model. \label{fig:swgp_r2_r4p4}}
\end{figure*}

\begin{figure*}
    \centering
        \includegraphics[width=0.97\textwidth]{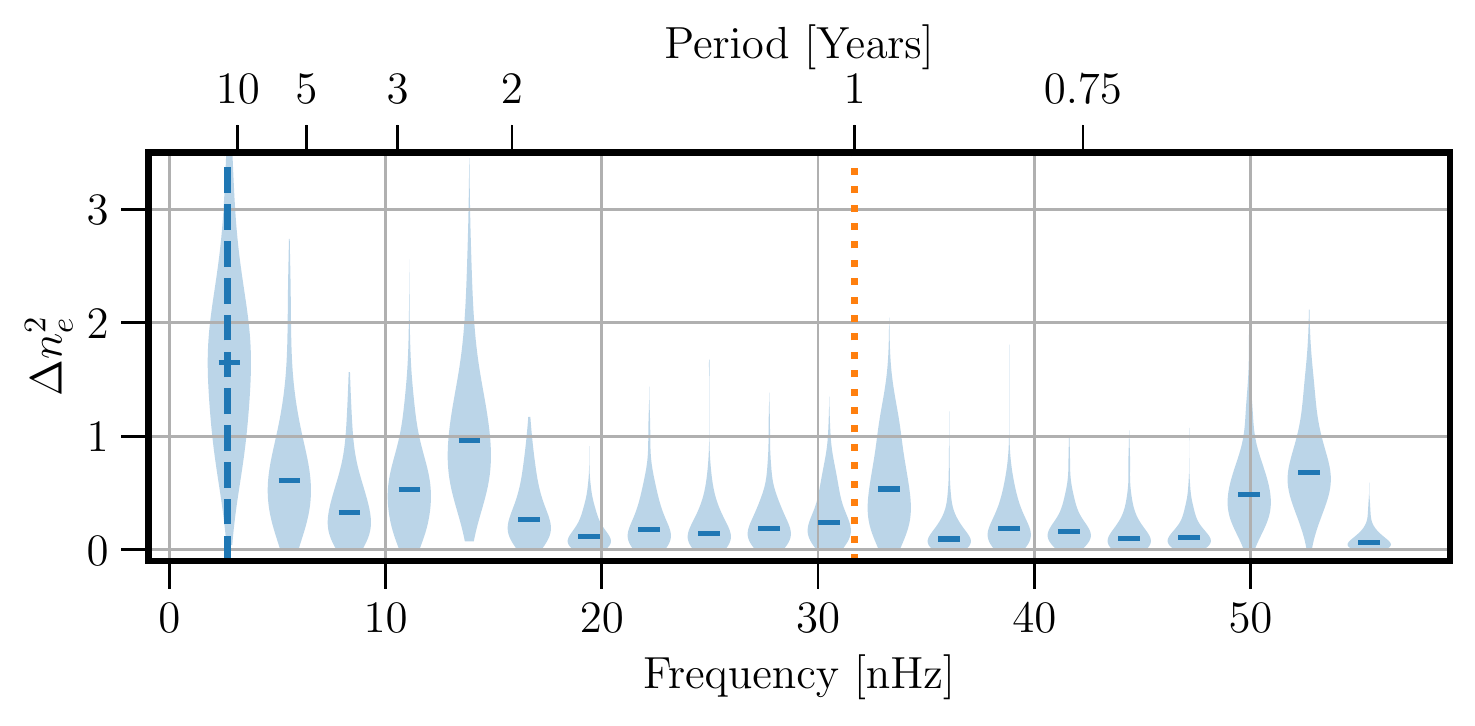}
        \caption{Power in the continuous solar wind density model. The violin plots show a simple representation of a power posterior calculated as $a^2_i+b^2_i$ for each set of Fourier coefficients. The horizontal bars show the median values. The orange dotted line shows a frequency of $1/{\rm yr}$ while the blue dashed vertical line shows $1/(11.75\,{\rm yr})$, an inverse average solar cycle. \label{fig:perturbation_power}}
\end{figure*}

In various test analyses we recorded the same type of behavior for these continuous models as was seen in \myfig{fig:3month_bin}, i.e., the fit returned large values of $n_E$ during periods of time when J0030+0451 and J1614$-$2230 observations had small values of $\theta_i$. Here we only report on the continuous model for $n_E$ where we also fitted for the same static, higher order $n^{(4.39)}_E$ term discussed in \mysec{subsec:sw_bins}. \myfig{fig:swgp_r2_r4p4} shows the resulting recovery of $n_E(t) = n^{mean}_E + n^{F}_E(t)$ from an analysis that uses the model in \myeq{eqn:n_fourier}.
The use of the Fourier basis gives us immediate access to an analogue of the power spectral density for the perturbations to the solar electron density. \myfig{fig:perturbation_power} shows violin plots for the power in the various frequencies calculated using the posteriors of the Fourier decomposition coefficients, $P=a_i^2+b_i^2$. Notice that a majority of the frequencies show small amounts of power, while the lowest frequency, $1/(11.4 {\rm \, yr})$, recovers the most power \textemdash\ very close to an inverse average solar cycle length of $11.75 \, {\rm yr}$. It will be interesting to see which frequencies have the most power in longer datasets.

A continuous signal like this would be a good candidate for a Gaussian process, however current code bases for PTAs construct different realizations of a process for each pulsar in an array. Since this work reveals a great deal of potential for gleaning information about the solar wind from PTAs we plan on developing per LOS solar wind perturbations for these pulsars in future work by using a simple deterministic inhomogeneous model. 

\section{Discussion}\label{sec:dis}

This work demonstrates a new set of Bayesian tools that allow for the use of PTA datasets, or any collection of pulsar timing data, as probes of the solar wind. The main innovations include a completely general Bayesian fit that includes timing model marginalization, generic spherically symmetric models for the solar electron density, and time dependent models. These models can be assembled in myriad ways in order to mitigate the solar wind as a noise source in pulsar data, or to use these sensitive datasets to study the Sun's coronal behavior across the solar system. The NANOGrav data set is not aimed at monitoring the solar wind, nonetheless it is sensitive to interesting solar phenomena. The tools we have developed, if applied to a PTA data set built for the purposes of solar science, have great potential.

One important aspect of this work not yet discussed is how these models can enable more in depth generic ISM studies. These solar wind models allow the removal of the DM variation signal from the local interplanetary medium in a principled way. This would better enable the use of these pulsar timing datasets for studying the ISM unencumbered by local variations in electron density.

The solar wind signal has been known to cause issues \citep{lommen+2006,splaver+2005} when trying to measure some timing effects in pulsar data sets---for instance, timing parallax, critical for constraining pulsar distances. The covariances with measuring parallax imply that improvements in the solar wind model could ultimately improve single source GW searches, as the distance to pulsars is an important aspect of pulsar term searches for GW from supermassive black binaries \citep{aab+19}. The solar wind has also been implicated as one of the systematics in the ``triple system'' \citep{archibald+2018}, impeding better limits of Einstein's equivalence principle. Better modeling of the solar wind might allow for more accurate measurements of these limits, in addition to allowing for better noise mitigation when these data are used for GW searches. Improvements to the handling of the solar wind in pulsar timing investigations stand to improve precision pulsar science in a wide variety of ways.

PTA studies of the solar wind have the potential to add a completely independent set of measurements to the myriad space missions studying solar weather. PTA observations survey a much wider swath of the Sun's environment than any spacecraft would be able to, taking column density information in 70+ lines of sight every month, probing beyond the outermost reaches of the solar environment. The observational campaigns of lower frequency observatories like the Canadian Hydrogen Intensity Mapping Experiment \citep{chime/pulsar2021}, LOFAR \citep{lofar2011,lofar2013} and the Long Wavelength Array \citep[LWA,][]{lwa_pulsar2015} are continually adding data that is extremely useful for DM variability studies, since these lower frequencies allow for more accurate measurement of the variations. These data can add important tests to the well developed solar weather modeling efforts undertaken by the solar physics community \citep{awsom2014}. These large scale measurements would allow for long timescale monitoring of the fast and slow solar wind, giving access to continuous measurements of the solar wind at all solar latitudes. 

\acknowledgments
{\it Author contributions.} This paper is the result of the work of many people and uses data from over a decade of pulsar timing observations.
We list specific contributions below. JSH developed and ran the solar wind analyses and led the paper writing. JS, and DRM contributed substantially to paper writing, discussion and interpretation of results. ZA, KC, MED, PBD, TD, JAE, RDF, ECF,  EF, PAG, GJ, MLJ, MTL, LL, DRL, RSL, MAM, CN, DJN, TTP, SMR, PSR, RS, IHS, KS, JKS, and WZ ran observations and developed the 11-year data set. 

{\it Acknowledgments.} The NANOGrav project receives support from National Science Foundation (NSF) as a Physics Frontier Center award number 1430284. We thank Caterina Tiburzi, Aditya Parthasarathy, Joseph Lazio and Patrick O'Neill for comments on an early draft of this manuscript. TD and MTL are supported by an NSF Astronomy and Astrophysics Grant (AAG) award number 2009468. ECF is supported by NASA under award number 80GSFC17M0002. The Dunlap Institute is funded by an endowment established by the David Dunlap family and the University of Toronto. TTP acknowledges support from the MTA-ELTE Extragalactic Astrophysics Research Group, funded by the Hungarian Academy of Sciences (Magyar Tudom\'anyos Akad\'emia), which was used during the development of this research. SMR is a CIFAR Fellow. Portions of this work performed at NRL were supported by ONR 6.1 basic research funding. Pulsar research at UBC is supported by an NSERC Discovery Grant and by CIFAR. WWZ is supported by the CAS Pioneer Hundred Talents Program and the Strategic Priority Research Program of the Chinese Academy of Sciences, grant No. XDB23000000.

\facilities{AO, GBO}

\software{Numpy \citep{numpy},
Scipy \citep{scipy},
MatPlotLib \citep{matplotlib},
\pint \citep{luo+2021},
\enterprise \citep{enterprise},
\extensions \citep{e_e},
\tempotwo \citep{tempo2},
\texttt{PTMCMCSampler} \citep{ptmcmc}
}

\appendix
\section{Generic Spherically Symmetric DM for Integer Powers}\label{sec:highpowers}
The expression in \myeq{eq:dm_rp} simplifies considerably for integer values of $p>1$. Here we show the first few expressions for powers greater than $p=2$ for reference.
\begin{eqnarray}
\text{DM}_{\odot}^{3}&=&n_{E}^{\left(3\right)}\left(1\,\text{au}\right)^{3}\frac{1+\cos\theta_{i}}{R_{E}^{2}\sin^{2}\theta_{i}}\\
\text{DM}_{\odot}^{4}&=&n_{E}^{\left(4\right)}\left(1\,\text{au}\right)^{4}\frac{2\pi-2\theta_{i}+\sin2\theta_{i}}{4R_{E}^{3}\sin^{3}\theta_{i}}\\
\text{DM}_{\odot}^{5}&=&n_{E}^{\left(5\right)}\left(1\,\text{au}\right)^{5}\frac{2-\cos\theta_{i}}{12R_{E}^{4}\sin^{4}\frac{\theta_{i}}{2}}
\end{eqnarray}

\section{Coronagraph Images}\label{sec:cme}
\begin{figure}[h]
\gridline{\fig{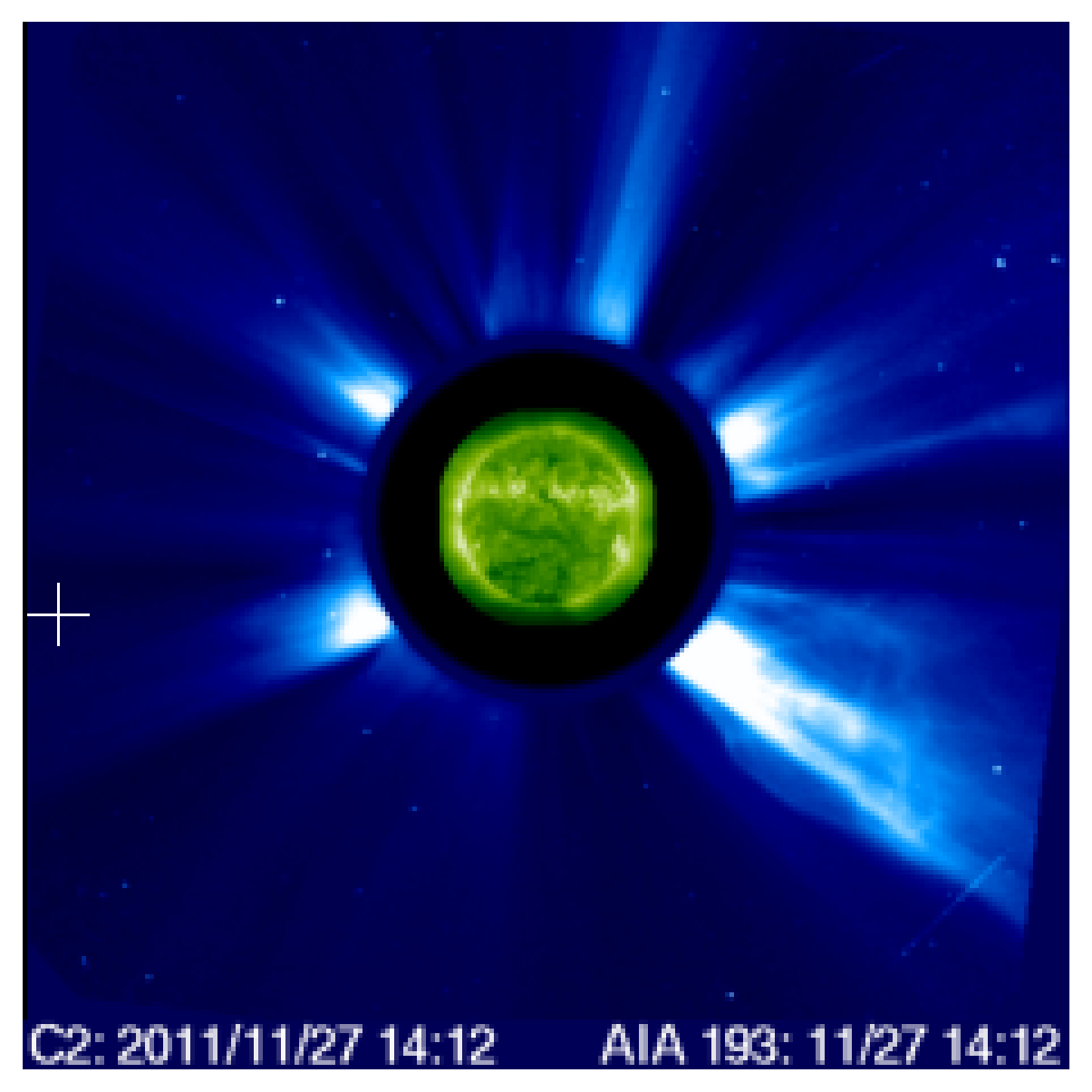}{0.3\textwidth}{(a)}
              \fig{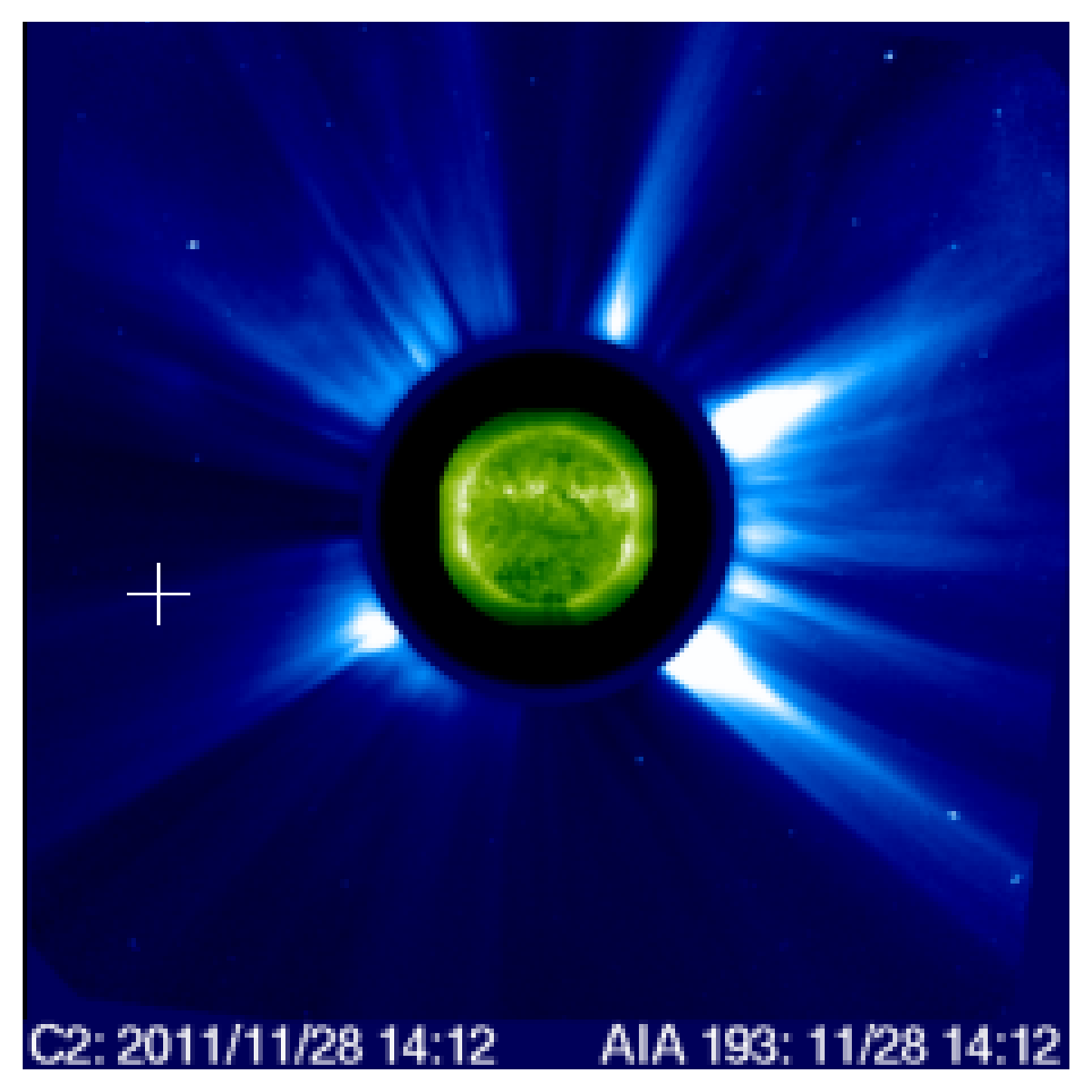}{0.3\textwidth}{(b)}
              \fig{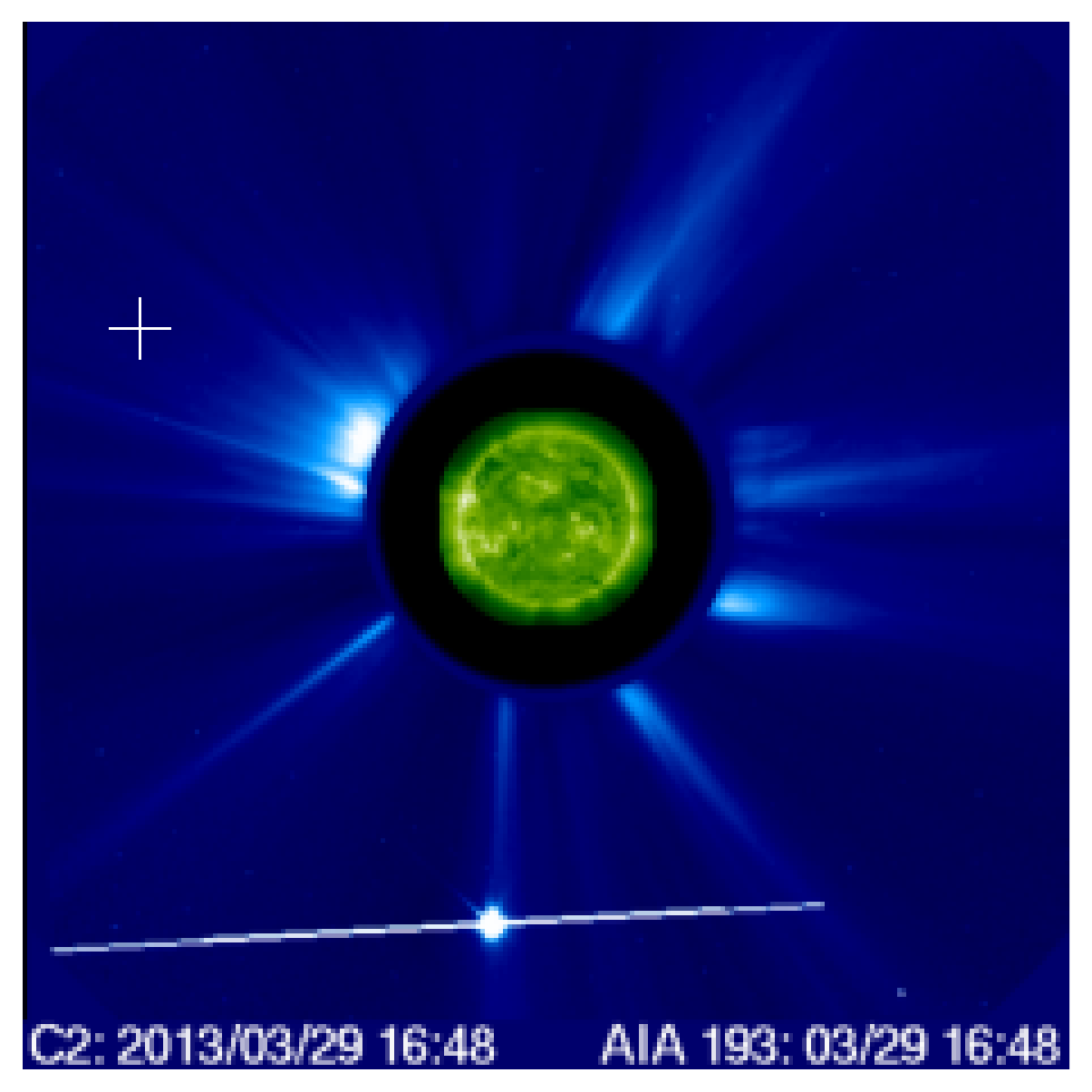}{0.3\textwidth}{(c)}}
\caption{Coronagraph images from the LASCO/C2 instrument onboard the SOHO spacecraft. The inset image is from the Extreme-Ultraviolet Imaging Telescope \citep[EIT,][]{eit:1995}. These coincide with the three pulsar timing observations discussed in \mysec{subsec:sw_bins}. The approximate pulsar positions are marked in each image with a white crosshairs. Images a) and b) are for the PSR J1614$-$2230 observations and are separated by one day. Image c) is for the PSR J0030+0451 observation. The dates and time in UT are given in the images.\label{fig:coronagraph}}
\end{figure}

As discussed in \mysec{subsec:sw_bins}, three observation epochs from \citetalias{abb+18a} have fairly small solar impact angles. In \myfig{fig:coronagraph} we show images from the Large Angle Spectroscopic Coronagraph \citep[LASCO,][]{lasco:1995} on board the Solar and Heliospheric Observatory (SOHO) from the same time as the pulsar observations, with the approximate pulsar positions marked. These white light images from the C2 camera can be related to the solar electron density \citep{ql:2002}, however such calculations are out of the scope of the current work.
These images are provided to highlight how close to the Sun these observations were taken ($<6R_\odot$) and how inhomogeneous the inner solar system can be. Testing various estimates of 2 and 3-dimensional electron density from these white light images would be an interesting use of dedicated pulsar observations close to the Sun.

\bibliography{sw}{}
\bibliographystyle{aasjournal}



\end{document}